\documentclass[graybox]{svmult}

\usepackage{mathptmx}       
\usepackage{helvet}         
\usepackage{courier}        
\usepackage{type1cm}        
\usepackage{makeidx}         
\usepackage{graphicx}        
\usepackage{multicol}        
\usepackage[bottom]{footmisc}
\usepackage{amsmath}
\usepackage{units}

\makeindex             

\newcommand{\super}[1]{^{\rm #1}}

\begin{document}

\title*{Simulating bicycle traffic by the Intelligent-Driver Model --
  reproducing the traffic-wave characteristics observed in a
  bicycle-following experiment}
\titlerunning{Simulating bicycle traffic by the Intelligent-Driver Model}
\author{Valentina Kurtc and Martin Treiber}

\institute{Valentina Kurtc \at Peter the Great St. Petersburg Polytechnic University, Polytechnicheskaya, 29, St. Petersburg, Russia, \email{kurtsvv@gmail.com}
\and Martin Treiber \at Technische Universit\"at Dresden, Institute for Transport and Economics, W\"urzburger Str. 35, 01062 Dresden, Germany \email{mail@martin-treiber.de}}
%
%
\maketitle

\abstract{Bicycle traffic operations become increasingly
  important and yet are largely ignored in the traffic flow community,
  until recently. We hypothesize that there is no qualitative
  difference between vehicular and bicycle traffic flow dynamics, so
  the latter can be described by reparameterized car-following models. To test this
  proposition, we reproduce bicycle experiments on a ring with the
  Intelligent-Driver Model and compare its fit quality (calibration) and predictive power (validation) with that of the Necessary-Deceleration-Model which is specifically designed for bike traffic. We find similar quality metrics for both models, so the above hypothesis of a qualitative equivalence cannot be rejected.}

\section{Introduction}
\label{sec:1}

In spite of its growing relevance, past research on bicycle traffic operations in
experiments~\cite{Navin1994bikeExperiments,Taylor1999bikeOperationsReview,Andresen2014_universal,Jiang2016traffic}
and models~\cite{Gould2009bikeCA,Andresen2014_NDM,Jin2015improved} is remarkably
scarce. In contrast, there is a multitude of empirical and
experimental
investigations for vehicular traffic flow, as well as a plethora of
models (for an overview see,
e.g.,~\cite{TreiberKesting-Book}). Therefore, it is natural to ask whether
there is a significant qualitative difference between vehicular and
bicycle traffic flow at all. In other words, the question arises if
one can use the well-developed car-following models for the simulation of bicycle
traffic instead of creating new bicycle
models.

In this paper, we test the Intelligent Driver Models
(IDM)~\cite{Treiber2000_PRE} as a typical
representative of car-following models against the "ring-road" bicycle traffic
experiments of Erik Andresen
et. al. \cite{Andresen2014_NDM,Andresen2014_universal}. In addition,
we compare the IDM fit quality with that of a specifically designed
``bicycle-following model'', the
Necessary-Deceleration-Model (NDM)~\cite{Andresen2014_NDM}.

In the following two sections, we shortly describe the models and the
experiments. Section~\ref{sec:4} specifies the calibration procedure
before we present our main calibration and validation results in
Sec.~\ref{sec:5} and conclude with a discussion in Sec.~\ref{sec:6}.

\section{Models under Investigation}
\label{sec:2}
Two microscopic car-following models are considered -- the IDM \cite{Treiber2000_PRE} and the NDM \cite{Andresen2014_NDM}. Both of them are formulated as coupled ordinary differential equations and characterized by an acceleration function which depends on the actual speed $v(t)$, the approaching rate $\Delta v(t )=v-v_l$ to the leader, and the gap $s(t)$.

The IDM is defined by the acceleration function \cite{Treiber2000_PRE}
\begin{equation}
	\dot{v}_{\rm IDM}(v,\Delta v,s)=a\left[1-\left(\frac{v}{v_0}\right)^4-\left(\frac{s^*(v,\Delta v)}{s}\right)^2\right],
\label{eq:01}
\end{equation}
where $s^*(v,\Delta v)=s_0+\max(0,vT+v \, \Delta
v/(2\sqrt{ab}))$ is the dynamically desired gap. The IDM contains five parameters to identify via
calibration -- $a,v_0,s_0,T,b$. Recently, it has been
found that stochasticity plays a
significant role for low-speed traffic flow, so we have added the simplest
form of white acceleration noise to the acceleration equation
(see~\cite{Treiber2017stochIDM_TRB} for details) when simulating
collective effects, cf. Sec.~\ref{subsec:stopGo}.

The NDM is originally formulated in terms of difference equations for
the speed~\cite{Andresen2014_NDM} which, in the limit of 
update times
$\Delta t$ tending to zero, is equivalent to a coupled differential equation with the
acceleration function
\begin{equation}
	\dot{v}_{\rm NDM}(v,\Delta v,s)=acc-\min(dec_1+dec_2,b_{\max}),
\label{eq:02}
\end{equation}
where
\begin{equation}
	acc =
 \begin{cases}
   0 &\text{$s \leq d(v)$}\\
    \frac{v_0-v}{\tau} &\text{$s>d(v)$},
 \end{cases}
 \label{eq:03}
\end{equation}
\begin{equation}
 	dec_1 = \min\left(\frac{(\Delta v)^2}{2(s-l-s_0)},b_{\max}\right),\	
\label{eq:04}
\end{equation}
\begin{equation}
		dec_2 =
\begin{cases}
  b_{\max}\frac{(s-d(v))^2}{(l-d(v))^2} &\text{$s \leq d(v), \ \Delta v \leq \epsilon$}\\
  0 &\text{otherwise.}
\end{cases}
\label{eq:05}
\end{equation}
The safety distance $d(v)=s_0+l+vT$ is a linear function of the cyclist's speed $v$, $l$ is a length of the cyclist. The NDM has 5 parameters to calibrate -- $\tau,v_0,s_0,T,b_{\max}$.

\section{Ring-road Experiment}
\label{sec:3}
Trajectory data of bicycle experiments were considered for calibration
and validation. These experiments were conducted by the University of
Wuppertal in cooperation with J\"ulich Forschungszentrum on 6th of May,
2012 \cite{Andresen2014_NDM}. Cyclists were moving one after another
along the oval track of 86 m length. However the measuring area
covered only a straight line of 20 m length. Group experiments were
performed for several density levels -- 5, 7, 10, 18, 20 and 33
participants.

Recently, further experiments with up to 63 cyclists have been
performed~\cite{Jiang2016traffic} giving essentially the same results and showing even more
pronounced stop-and-go waves for the higher densities.

\section{Methods}
\label{sec:4}
%
We have estimated the model performance by two approaches. One is based on
trajectories, and the other on aggregated properties of scatter plots derived
from stationary measurements (virtual detectors).

\subsection{Calibrating and Validating Trajectories}
\label{subsec:calMeth}
Pairs of consecutive trajectories were used for calibration and
  validation according to the global approach
\cite{TreiberKesting-Book,Kurtc2016}. Specifically, the
microscopic model was initialized with the empirically given speed
$v\super{sim}(t=0)=v\super{data}(t=0)$ and gap
$s\super{sim}(t=0)=s\super{data}(t=0)$, and the trajectory of the
following cyclist  for a given leader trajectory was calculated using the ballistic
update (see Chapter 10.2 of ~\cite{TreiberKesting-Book}) with a
time step $\Delta t=0.04 s$. Afterwards, the simulated gaps
  $s\super{data}(t)$ were compared with the experimentally observed
  gaps by means of two objective functions, namely the absolute error measure
\begin{equation}
	S\super{abs}=\frac{\sum_{i=1}^n(s_i\super{sim}-s_i\super{data})^2}{\sum_{i=1}^n(s_i\super{data})^2},
\label{eq:07}
\end{equation}
and the relative error measure
\begin{equation}
	S\super{rel}=\frac{1}{n}\sum_{i=1}^n\left(\frac{s_i\super{sim}-s_i\super{data}}{s_i\super{data}}\right)^2.
\label{eq:08}
\end{equation}

\subsection{Comparing Microscopic Fundamental Diagrams}
\label{subsec:compareFD}
We used virtual stationary detectors at several positions of the ring and calculated
\begin{itemize}
	\item{the instantaneous speed $v_i$ of cyclist $i$ at passage time,}
	\item{the "microscopic density" $\rho_i$, i.e., the inverse space headway to the leader at passage time,}
	\item{and the microscopic flow $Q_i=\rho_iv_i$, i.e., the inverse (time) headway.}
\end{itemize}

We combined these data to microscopic speed-density scatter plots
both for the simulations and the experiments by defining the sets
$\{[\rho_i\super{data},v_i\super{data}]\}$ and
$\{[\rho_i\super{model},v_i\super{model}]\}$,
respectively. In order to quantitatively compare the similarity, we create a partitioning of the data points by filtering them according to $N$ equally-spaced density intervals,
\begin{equation}
	V_j\super{src}=\{v_i^{(src)}:\rho_i^{(src)} \in [\rho_j,\rho_{j+1}],i=1,...,M_{src}\},
\label{eq:09}
\end{equation}
where $j=1,...,N, \text{src}=\{\text{data,model}\}$. 
The key idea is
to interpret $V_j\super{src}$ as a one-dimensional random variable. Now
we calculate the (cumulative) distribution functions both for the experiment,
$F_{V_j\super{data}}(x)$, and the simulations, $F_{V_j\super{model}}(x)$,
and calculate the Kolmogorov-Smirnov distance
\begin{equation}
	D_j=\sup_{x}|F_{V_j\super{data}}(x)-F_{V_j\super{model}}(x)|,j=1,...,N
\label{eq:10}
\end{equation}
The distance $D_j$ averaged over all density bins provides the quantitative metric which estimates the similarity of two data-clouds
\begin{equation}
	D^*=\frac{1}{N}\sum_{j=1}^N D_j
\label{eq:11}
\end{equation}

\section{Results}
\label{sec:5}

\begin{figure}[h]
\centering{\includegraphics[scale=0.6]{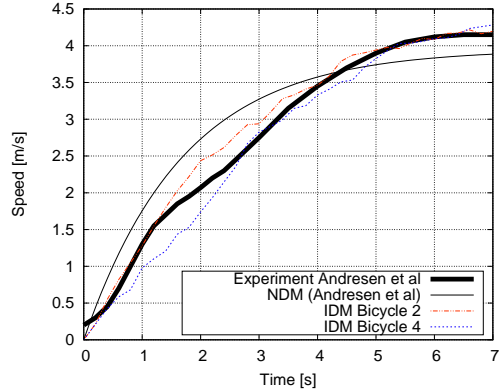}}
\caption{First stage of the free acceleration of cyclists. Shown are
  a typical experimental profile and the NDM prediction taken
  from~\cite{Andresen2014_NDM}, and two realisations from the
  stochastic IDM}
\label{fig:free}       
\end{figure}

\subsection{Free acceleration}
\label{subsec:calFree}
First, we compare and calibrate the free acceleration profile against
the experimental results (cf. Fig.~4
of~\cite{Andresen2014_NDM}). After calibrating the relevant IDM
parameters to the values $v_0=\unit[4.3]{m/s^2}$ and
$a=\unit[1.0]{m/s^2}$ and adding a small amaount of white noise
(intensity $Q=\unit[0.02]{m^2/s^3}$), we found
a better agreement compared to the NDM (Fig.~\ref{fig:free}).

\subsection{Collective Driving Behavior}
\label{subsec:calCong}
Both models have been calibrated for all trajectory pairs and optimal parameter value distributions were obtained. Table \ref{tab:01} presents the calibration errors. For both error measures (Eq. \ref{eq:07} and \ref{eq:08}) lower error values correspond to the IDM whereas higher errors come from the NDM.

The use of several error measures can be interpreted as a benchmark
for the robustness of the model calibration. Specifically, for a
  good model,  the calibration results and the distribution of the
  calibrated parameters should not significantly vary with the chosen
  error measure. We compare the Kolmogorov-Smirnov distance
\begin{equation}
	D_{n}=\sup\left|F_{1,n}(x)-F_{2,n}(x)\right|
\label{eq:12}
\end{equation}
of the distributions $F_{1,n}$ and $F_{2,n}$ of parameter $n$ as
obtained by calibrating the 
trajectories with respect to the absolute and relative error measure
$S\super{abs}$ and $S\super{rel}$,
respectively. According to the results presented in Table
\ref{tab:02}, the IDM tends to be slightly more robust than the
NDM.

\begin{table}
  \quad
  \begin{minipage}[b]{35mm}
\caption{Calibration errors (\%) for IDM and NDM}
\label{tab:01} 
\begin{tabular}{p{1cm}p{1cm}p{1cm}}
\svhline\noalign{\smallskip}
Model & $\sqrt{S\super{abs}}$ & $\sqrt{S\super{rel}}$  \\
\noalign{\smallskip}\hline\noalign{\smallskip}
IDM & 2.86  & 3.01\\
NDM & 4.85 & 5.05\\
\noalign{\smallskip}\svhline\noalign{\smallskip}
\end{tabular}

\end{minipage}
\qquad
\begin{minipage}[b]{60mm}
\caption{Kolmogorov-Smirnov distance (\ref{eq:12}) $D_p$ of the parameter values for the two models}
\label{tab:02}
\begin{tabular}{p{10mm}p{10mm}p{10mm}p{10mm}p{10mm}p{10mm}}
\svhline\noalign{\smallskip}
IDM & $D_a$ & $D_{v_0}$ & $D_{s_0}$ & $D_T$  \\
\noalign{\smallskip}\hline\noalign{\smallskip}
& 0.023 & 0.027 & 0.036 & 0.029\\
\noalign{\smallskip}\svhline\noalign{\smallskip}
NDM & $D_\tau$ & $D_{v_0}$ & $D_{s_0}$ & $D_T$ & $D_{b_{\max}}$  \\
\noalign{\smallskip}\hline\noalign{\smallskip}
& 0.043 & 0.035 & 0.072 & 0.026 & 0.039\\
\noalign{\smallskip}\svhline\noalign{\smallskip}
\end{tabular}

\end{minipage}
\end{table}

\begin{figure}[h]
\centering{\includegraphics[scale=0.8]{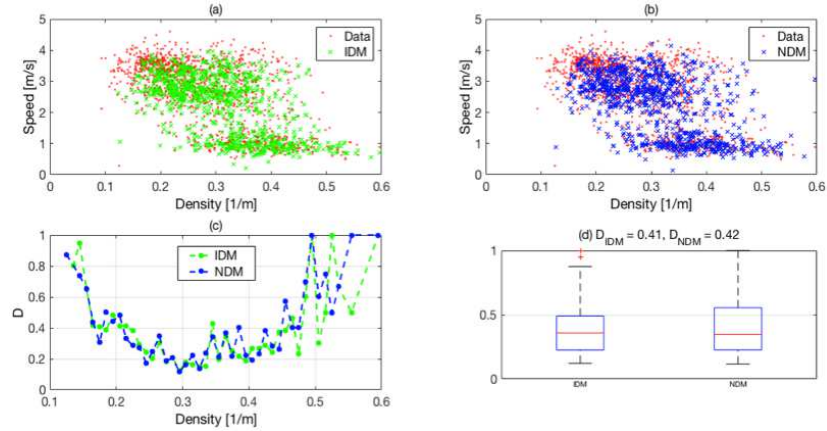}}
\caption{Speed-density relation for the data, the IDM (a) and the NDM (b), (c) -- values of the metric $D_j$ for $j$th density bin, (d) -- boxplots corresponding to the IDM (left) and the NDM (right)}
\label{fig:1}       
\end{figure}

\subsection{Microscopic Fundamental Diagrams Comparison}
\label{subsec:microFD}
We have calculated the microscopic fundamental diagram and the distance measures both for the
real data and for the simulation of the two models with the
optimal parameter values without noise. The results are shown in Fig.~\ref{fig:1}.

\begin{figure}[h]
\centering{\includegraphics[scale=0.65]{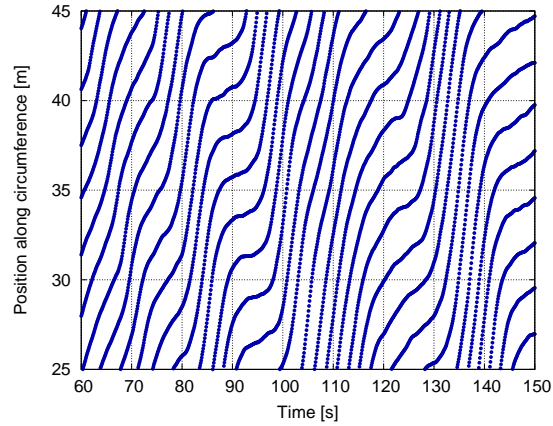}}
\caption{Simulated trajectories of stop-and-go traffic appearing for
  the IDM for dense traffic (density 300 cyclists/km)}
\label{fig:stopGo}       
\end{figure}

\subsection{Stop-and-go waves}
\label{subsec:stopGo}
%
Besides calibrating the IDM by trajectory pair, we
also tested if the IDM can produce collective effects such as the stop-and-go traffic observed in
the experiments~\cite{Andresen2014_NDM}
and~\cite{Jiang2016traffic}. Figure~\ref{fig:stopGo} shows the
result. Instead of using heterogeneous drivers,
we simplified the investigation as much as possible by using a single
parameter set for all drivers replacing the heterogeneity by white
acceleration noise. In contrast to the free-flow simulation
(Fig.~\ref{fig:free}), a higher
amount of $\unit[0.1]{m^2/s^3}$ was needed to approximatively
reproduce the observed amplitude and frequency statistics of the
traffic waves while the free-flow parameters $v_0=\unit[4.3]{m/s^2}$ and
$a=\unit[1.0]{m/s^2}$ were the same. The calibrated values
$T=\unit[0.85]{s}$, $s_0=\unit[0.4]{m}$, and $b=\unit[1.3]{m/s^2}$
were near the median of the trajectory-by-trajectory calibration of Section~\ref{subsec:calCong}.

\subsection{Inter-Driver Variation and Validation}
\label{subsec:val}
%
Validation by cross comparison implies determining the error
  measures for a certain test data set by simulating the model with the
parameters calibrated to the disjunct ``learning'' data
set~\cite{TreiberKesting-Book}.  For each experiment (5,
10, 15, 18, 20 and 33 participants, respectively), we have separately calculated the
calibration-validation matrix whose elements $M_{ij}$ give the
error measure $\sqrt{S\super{abs}}$ for the trajectory pair $j$ as obtained from the
model calibrated to the trajectory pair $i$. The diagonal element $M_{ii}$ are
the calibration errors whereas the off-diagonal elements $M_{ji}$, $j
\neq i$, give a superposition of the validation error and the
inter-driver variation of follower $j$ with respect to follower
$i$. The average validation error $\epsilon\super{val}$ and calibration error $\epsilon\super{cal}$ are given by
\begin{equation}
	\epsilon\super{val}=\frac{1}{n(n-1)} \sum_{i=1}^n \sum_{\substack{j=1 \\ j \neq i}}^n M_{ij},
\label{eq:13}
\end{equation}
\begin{equation}
	\epsilon\super{cal}=\frac{1}{n} \sum_{i=1}^n M_{ii},
\label{eq:14}
\end{equation}
where $n$ is a number of trajectory pairs.
Notice that a separation of these two causes would require using
disjunct parts of the same trajectory for calibration and validation
which is only viable for longer trajectories than the available ones.
To obtain measures for the overall fitting quality and the predictive
power plus inter-driver variations, we have calculated the ratio of the average
validation error to the calibration error (Table \ref{tab:03}). 

\begin{table}
\caption{Calibration, validation errors (\%) and averaged ratios for IDM and NDM}
\label{tab:03}       
%
%
\begin{center}
\begin{tabular}{p{2.3cm}p{1.3cm}p{1.3cm}p{1.3cm}p{1.3cm}p{1.3cm}p{1.3cm}}
\svhline\noalign{\smallskip}
 & $N=5$ & $N=10$ & $N=15$ & $N=18$ & $N=20$ & $N=33$  \\
\noalign{\smallskip}\hline\noalign{\smallskip}
IDM &  &  &  &  &  &   \\
\noalign{\smallskip}\hline\noalign{\smallskip}
Calibration error & 1.74  & 3.08 & 2.87  & 11.98 & 3.43  & 5.67\\
Validation error & 32.23 & 26.37 & 22.53  & 32.79 & 26.6  & 32.63\\
Ratio & 18.5 & 8.5 & 7.8  & 2.7 & 7.7  & 5.7\\
\noalign{\smallskip}\hline\noalign{\smallskip}
NDM &  &  &  &  &  &   \\
\noalign{\smallskip}\hline\noalign{\smallskip}
Calibration error & 1.28  & 6.39 & 3.49  & 3.54 & 5.78  & 7.43\\
Validation error & 42.58 & 32.51 & 29.19  & 35.58 & 31.67  & 27.58\\
Ratio & 33.2 & 5.1 & 8.3  & 10.0 & 5.5  & 3.7\\
\noalign{\smallskip}\svhline\noalign{\smallskip}
\end{tabular}
\end{center}
\end{table}

\section{Discussion and Conclusions}
\label{sec:6}
According to the results presented in this paper, we conclude that the
IDM, which has a similar underlying heuristics as the NDM, can not
only describe vehicular but also bicycle traffic, or, at least,
"bicycle following".The IDM trajectories fit even better to the data
than that of the NDM. The IDM calibration errors with absolute and
relative error measures are 2.86 and 3.01 \% whereas for the NDM they
are 4.85 and 5.05 \% respectively. The application of several
objective functions indicates that the IDM calibration is also
more robust in comparison to the NDM. Validation results show that the
predictive power of the IDM is better than that of the NDM. However,
the validation results are confounded by discrepancies from
inter-driver variations, so further investigations to separate these
factors are necessary. 

The analysis of macroscopic characteristics such as speed-density
relations provides more or less the same results for both
models. Specifically, the averaged Kolmogorov distance $D^*$
(Eq. \ref{eq:11}) is nearly the same. Furthermore,
the stochastic IDM can well describe the statistical features of
the amplitude and frequency of the observed
stop-and-go waves. We conclude that the
dynamics of bicycle traffic differs only quantitatively from vehicular
traffic and reparameterized car-following models such as the IDM work at least as well
as dedicated ``bike-following'' models.

\bibliographystyle{spmpsci}
\bibliography{tgffinalbibs}

\end{document}